\documentstyle[aps,preprint]{revtex}
\begin{document}
\title{Photon-neutrino interactions in magnetic fields}
\author{R.~ SHAISULTANOV\thanks{%
Email:shaisultanov@inp.nsk.su}}
\address{Budker Institute of Nuclear Physics \\
630090, Novosibirsk 90, Russia}
\maketitle

\begin{abstract}
The low-energy two neutrino-two photon interactions in the presence of
homogeneous magnetic field are studied. The cross sections in external
magnetic field are shown to be larger than in vacuum by factor $\sim \left(
m_W\ /m_e\right) ^4\left( B/B_c\right) ^2$. The energy-loss rate due to the
process $\gamma \gamma \rightarrow \nu \overline{\nu }$ in magnetic field is
obtained.
\end{abstract}


\newpage
\narrowtext

Low-energy neutrino-photon interactions may be of interest in astrophysics
and cosmology. A well-known example is the neutrino pair production via $%
\gamma \ \gamma \rightarrow \nu \ \overline{\nu }$ , which provides an
energy loss mechanism for stellar processes, and related processes, such as
the neutrino photon scattering $\gamma \ \nu \rightarrow \gamma \ \nu \ $
,that also may be important in the study of stellar evolution. Unfortunately
the amplitude in any channel is known to be highly suppressed . In \cite{a}
Gell-Mann showed that in the four-Fermi limit of the standard model the
amplitude is exactly zero to order $G_F$ because by Yang theorem \cite{yang}
two photons cannot couple to $J=1$ state. Therefore the amplitude is
suppressed by additional factors of $\omega /m_W$ , where $\omega $ is the
photon energy and $m_W$ is the $W$ mass \cite{a,a2,b,c}. For example, in the
case of massless neutrinos, the amplitude for $\gamma \nu \rightarrow \gamma
\nu $ in the Standard model is suppressed by factor $1/m_W^2$ \ \cite{f} .
As a result the cross section is exceedingly small. Also recently Dicus and
Repko \cite{f1} considered the processes $\gamma \nu \rightarrow \gamma
\gamma \nu $, $\gamma \gamma \rightarrow \gamma \nu \bar{\nu}$ and $\nu \bar{%
\nu}\rightarrow \gamma \gamma \gamma $ , the cross sections are shown to be
much larger than the elastic cross section $\sigma (\gamma \nu \rightarrow
\gamma \nu )$ for photon energies $\omega >1$\thinspace keV. By now it is
well known that in many astrophysical environments the absorption, emission,
or scattering of neutrinos and photons occurs in the presence of strong
magnetic fields with strength of order $B_{{\rm c}}=m_e^2/e=4.41\times
10^{13}~{\rm Gauss}$ \cite{Raffelt}. Many processes were studied in the
presence magnetic fields. Among them, for example, photon splitting $\gamma
\to \gamma \gamma $ \cite{split} and neutrino Cherenkov process $\nu \to \nu
\gamma $ \cite{wun}.

In this paper we will consider low-energy two neutrino-two photon
interactions in the presence of homogeneous magnetic field. At $B/B_c\leq
0.1 $ it is enough to use following effective action \cite{f1}: 
\begin{equation}
{\cal L}_{{\rm eff}}=4\frac{G_F\,a}{\sqrt{\displaystyle 2}}\frac{\alpha
^{3/2}}{\sqrt{\displaystyle 4\pi }}\frac 1{m_e^4}\left[ \frac 5{180}\left(
N_{\mu \nu }F_{\mu \nu }\right) \left( F_{\lambda \rho }F_{\lambda \rho
}\right) -\frac{14}{180}N_{\mu \nu }F_{\nu \lambda }F_{\lambda \rho }F_{\rho
\mu }\right]  \label{1}
\end{equation}

where $N_{\mu \nu }$ is 
\begin{equation}
N_{\mu \nu }=\partial _\mu \left( \bar{\psi}\gamma _\nu (1+\gamma _5)\psi
\right) -\partial _\nu \left( \bar{\psi}\gamma _\mu (1+\gamma _5)\psi
\right) \,,
\end{equation}
and $a=1-$ $\frac 12\left( 1-4\sin ^2\theta _W\right) $ for $\nu _e$ , where
the first term in $a$ is the contribution from the W exchange diagram and
second one from the Z exchange diagram \cite{f1}. For $\nu _\mu $ and $\nu
_\tau $ the W  contribution is multiplied by very small factor $\left(
m_e/m_\mu \right) ^4$ or $\left( m_e/m_\tau \right) ^4$ and thus we can
ignore it. Now taking into account only the Z exchange contribution we will
get numerically small $a\simeq -\frac 12+2\sin ^2\theta _W=-0.04$ and hence
in the subsequent discussion we will deal only with electron neutrinos.
Notice also that the effective Lagrangian (\ref{1}) provides an adequate
description of processes with $\omega <m_e$ \cite{f1}.In this action we now
substitute $F_{\mu \nu }\rightarrow F_{\mu \nu }+\ f_{\mu \nu }\ ,$ where $\
f_{\mu \nu }$ is quantized electromagnetic field and $F_{\mu \nu }$ from now
on denotes external field. Retaining terms bilinear in $\ f_{\mu \nu }$ we
obtain final effective action

\begin{eqnarray}
{\cal L}_{{\rm eff}} &=&4\frac{G_F\,a}{\sqrt{\displaystyle 2}}\frac{\alpha
^{3/2}}{\sqrt{\displaystyle 4\pi }}\frac 1{180m_e^4}N_{\mu \nu }\left[
5\left( F_{\mu \nu }\ f_{\lambda \rho }\ f^{\lambda \rho }+2\ f_{\mu \nu }\
\ f^{\lambda \rho }F_{\lambda \rho }\right) -\right.  \label{2} \\
&&\left. -14\left( F_{\nu \lambda }\ f^{\lambda \rho }\ f_{\rho \mu
}+F^{\lambda \rho }\ f_{\nu \lambda }\ f_{\rho \mu }+F_{\rho \mu }\
f^{\lambda \rho }\ f_{\nu \lambda }\right) \right]  \nonumber
\end{eqnarray}

Using (\ref{2}) we can find the amplitudes for $\gamma \gamma \rightarrow
\nu \overline{\nu }$ and cross related reactions. The amplitude for $\gamma
\gamma \rightarrow \nu \overline{\nu }$ can be represented in the form 
\begin{equation}
{\cal M}=8\frac{G_F\,a}{\sqrt{\displaystyle 2}}\frac{\alpha ^{3/2}}{\sqrt{%
\displaystyle 4\pi }}\frac 1{180m_e^4}\overline{u}\left( p_1\right) \gamma
_\mu \left( 1+\gamma _5\right) \upsilon \left( p_2\right) {\cal J}^\mu
\label{amp}
\end{equation}

where 
\begin{eqnarray}
{\cal J}^\mu &=&6\left( Fk_1\right) ^\mu \ \left[ \left( k_1\varepsilon
_2\right) \left( k_2\varepsilon _1\right) -\left( k_2k_1\right) \left(
\varepsilon _2\varepsilon _1\right) \right] -6k_2^\mu \ \left(
k_1F\varepsilon _1\right) \left( k_1\varepsilon _2\right) -  \label{ampl} \\
&&-14k_1^\mu \ \left[ \left( \varepsilon _2Fk_1\right) \left( k_2\varepsilon
_1\right) -\left( k_2Fk_1\right) \left( \varepsilon _2\varepsilon _1\right)
+\left( \varepsilon _1Fk_2\right) \left( k_1\varepsilon _2\right) -\left(
\varepsilon _1F\varepsilon _2\right) \left( k_2k_1\right) \right] + 
\nonumber \\
&&+6\varepsilon _2^\mu \ \left( k_1F\varepsilon _1\right) \left(
k_1k_2\right) +28\varepsilon _1^\mu \ \left[ \left( \varepsilon
_2Fk_1\right) \left( k_2k_1\right) -\left( k_2Fk_1\right) \left( \varepsilon
_2k_1\right) \right] +\left( k_1\leftrightarrow k_2,\varepsilon
_1\leftrightarrow \varepsilon _2\right)  \nonumber
\end{eqnarray}

The amplitudes for $\gamma \nu \rightarrow \gamma \nu $ and $\gamma 
\overline{\nu }\rightarrow \gamma \bar{\nu}$ can be easily obtained from (%
\ref{amp},\ref{ampl}) with the use of cross symmetry. Then the cross section
of process $\gamma \gamma \rightarrow \nu \overline{\nu }$, averaged over
polarizations of incoming photons, is given by

\begin{eqnarray}
\sigma _B(\gamma \gamma &\rightarrow &\nu \overline{\nu })=\frac 1{3(180)^2}%
\frac{G_F^{\,2}\,\alpha ^3a^2}{\pi ^2m_e^8}\left[ 1112\left( k_1k_2\right)
^2\ \left[ \left( k_1F^2k_1\right) +\left( k_2F^2k_2\right) \right] +\right.
\label{result} \\
&&\left. +1480\left( k_1k_2\right) ^2\ \left( k_2F^2k_1\right) -440\left(
k_1k_2\right) \ \left( k_2Fk_1\right) ^2+392\left( k_1k_2\right) ^3\ F_{\mu
\nu }F^{\mu \nu }\right]  \nonumber
\end{eqnarray}

From (\ref{result}) the cross sections of neutrino-photon interactions in
magnetic field $\sigma _B$ can be easily estimated. If we take, for example, 
$k_1=\omega $ $(1,1,0,0)$ and $k_2=\omega $ $(1,-1,0,0)$ with $%
\overrightarrow{B}=(0,0,B)$ we will get

\begin{equation}
\sigma _B=\ 6.6\cdot 10^{-51}\left( \frac \omega {m_e}\right) ^6\left( \frac 
B{B_c}\right) ^2cm^2\   \label{3}
\end{equation}

A comparison between (\ref{3}) and results of \cite{a2,b,c,f} shows that in
external magnetic field neutrino-photon cross sections are enhanced by
factor $\sim \left( m_W\ /m_e\right) ^4\left( B/B_c\right) ^2$.

Let us now study how the process $\gamma \gamma \rightarrow \nu \overline{%
\nu }$ contribute to stellar energy-loss of a star with strong magnetic
field $B=10^{12}-10^{13}~{\rm Gauss}$ ( neutron star ). The energy loss rate
( in ergs/sec cm$^3$ ) in our case is equal to

\begin{equation}
Q=\frac 1{\left( 2\pi \right) ^6}\int \frac{2d^3k_1}{e^{\omega _1/T}-1}\int 
\frac{2d^3k_2}{e^{\omega _2/T}-1}\frac{\left( k_1k_2\right) }{\omega
_1\omega _2}\left( \omega _1+\omega _2\right) \text{ }\sigma _B\left( \gamma 
\text{ }\gamma \rightarrow \nu \text{ }\overline{\nu }\right)  \label{qqq}
\end{equation}

After integration we obtain

\begin{eqnarray}
\text{ }Q &=&\frac{256}{637875}\left[ 973\pi ^2\text{ }\zeta \left( 5\right)
+12610\text{ }\zeta \left( 7\right) \right] \frac{G_F^2\text{ }\alpha ^2a^2}{%
m_e^4}\left( \frac B{B_c}\right) ^2\text{ }T^{13}=  \label{qqq1} \\
&=&0.4\cdot 10^{10}\text{ }T_9^{13}\text{ }\left( \frac B{B_c}\right) ^2%
\frac{erg}{s\text{ }cm^3}  \nonumber
\end{eqnarray}

where $T_{9\text{ }}$is temperature in units of $10^9$ $^{\circ }K$. A
comparison between Q and energy-loss rates due to other processes, such as
pair neutrino and photoneutrino processes ( see e.g.\cite{shab,dic72,bea} ),
shows that, at $T\geq 10^9$ $^{\circ }K$ , Q cannot be neglected in
astrophysical considerations. In addition taking into account the statement
of \cite{tep} that the family of processes $\gamma \nu \rightarrow \gamma
\gamma \nu $, $\gamma \gamma \rightarrow \gamma \nu \bar{\nu}$ and $\nu \bar{%
\nu}\rightarrow \gamma \gamma \gamma $ , whose cross sections are of order $%
\,\sim 10^{-55}$ $\left( \omega /m_e\right) ^{10}cm^2$ \cite{f1}, could be
quite important in astrophysics, we may conclude that processes discussed
above may also be of some importance in astrophysics.

\end{document}